\documentclass[final,5p,times,twocolumn]{elsarticle}
\usepackage{tikz,pstricks}
\usetikzlibrary{shapes,chains,arrows,positioning, calc, backgrounds}
\usepackage{amssymb,amsmath}
\usepackage{xcolor}
\usepackage{balance}
\usepackage{bm}

\tikzstyle{block} = [draw, shape=rectangle, minimum height=3em, minimum width=3em, node distance=2cm, line width=1pt]

\begin{document}

\begin{frontmatter}

\title{Flattening the curves: on-off lock-down strategies for COVID-19 with an application to Brazil}

\author[CEFETLabel]{Luís Tarrataca}
\author[FRURJ]{Claudia Mazza Dias}
\author[CEFETLabel]{Diego Barreto Haddad}
\author[COPPE,CU]{Edilson Fernandes De Arruda}

\address[CEFETLabel]{Department of Computer Engineering, Celso Suckow da Fonseca Federal Center for Technological Education, Brazil}
\address[FRURJ]{Department of Technologies and Languages Multidisciplinary Institute, Federal Rural University of Rio de Janeiro, Brazil}
\address[COPPE]{Alberto Luiz Coinbra Institute-Graduate School and Research in Engineering, Federal University of Rio de Janeiro, Brazil}
\address[CU]{School of Mathematics, Cardiff University, Senghennydd Rd, Cardiff CF24 4AG, UK}

\begin{abstract}
The current COVID-19 pandemic is affecting different countries in different ways. The assortment of reporting techniques alongside other issues, such as underreporting and budgetary constraints, makes predicting the spread and lethality of the virus a challenging task. This work attempts to gain a better understanding of how COVID-19 will affect one of the least studied countries, namely Brazil. Currently, several Brazilian states are in a state of lock-down. However, there is political pressure for this type of measures to be lifted. This work considers the impact that such a termination would have on how the virus evolves locally. This was done by extending the SEIR model with an on / off strategy. Given the simplicity of SEIR we also attempted to gain more insight by developing a neural regressor. We chose to employ features that current clinical studies have pinpointed has having a connection to the lethality of COVID-19. We discuss how this data can be processed in order to obtain a robust assessment.
\end{abstract}


\end{frontmatter}


\section{Introduction \label{sec:introduction}}


The Coronavirus Disease 2019, whose aetiological agent is known as Severe Acute Respiratory Syndrome Coronavirus 2 (SARS-CoV-2)~\cite{RodriguezClinical2020}, has been dubbed COVID-19 by the World Health Organization (WHO). The virus has been spreading worldwide and was effectively classified as a pandemic by the WHO~\cite{WHOReport51}. The first cases were reported to the chinese bureau of WHO in December 2019 in Wuhan City, Hubei Province of China~\cite{WHOReport1}. Given that the pandemic is still quite recent, several efforts are underway to try to predict its evolution, namely in terms of spread, infection rates, mortality, amongst other dimensions~\cite{ferguson2020,flaxman2020,walker2020, jia2020,lourenco2020}.

However, by checking interactive web-based dashboards (\textit{e.g.,}~\cite{dong2020a}) it has become clear that, globally, reporting methods appear to differ substantially from country to country. Possible factors for such divergences may include lack of testing facilities, monetary constraints, geographical scale, under-reporting and even political unwillingness to divulge the true scale. Given the reliability issues related to the reported data, there is no consensus over the mortality rates associated to COVID-19 (\textit{e.g., }\cite{Spychalski2020}). For example, while~\cite{Baud2020} argues that the rates are overestimated,~\cite{Wu2020-mort} argue otherwise. Consequently, some have started to question whether the COVID-19 epidemic can be managed on the basis of daily data~\cite{casella2020}.

Understandably, most of the studies have focused on the contagion scenarios in Europe and China. To our knowledge, there appears to be a lack of COVID-19 related research focusing on south America, more specifically Brazil, home of approximately 211 million people, the world's fifth-largest country by area and currently the world's $8^{th}$ largest economy. 
Brazil-specific predictions incorporating government introduced mitigation strategies were made available in~\cite{batista2020a} for the states of São Paulo and Rio de Janeiro. These represent the two largest economic units of the union and also concentrate a significant part of the population. However, the true local scale is difficult to assess. In part, this is due to under-reporting of cases owing to chronic tests shortages~\cite{batista2020b}. Furthermore, the official figures only include deaths reported by hospitals. A more detailed analysis of current research is presented in Section~\ref{sec:relatedWork}.

The set of guiding questions behind this work can be stated as follows: It is common knowledge that political leadership in Brazil has at times conveyed contradictory messages on how best to tackle the crisis. Some argue for the necessity of mitigation measures, whilst others defend that these will result in insurmountable damage to the economy. As a result, can public trust in civil servants affect the epidemic? Given the current set of Brazilian public policies aiming at mitigation, how will this affect the local spread of COVID-19? What would be the effects of more relaxed non-pharmaceutical measures? Section~\ref{sec:SEIRModelOnOffStrategy} exploits these questions by proposing a carefully designed quarantine strategies based on the availability of hospitalisation beds and evaluating these strategies in time by means of a traditional SEIR epidemic model. 

Furthermore, is it possible to predict how COVID-19 will affect Brazil based on what is happening in other countries? What features should be considered? Are there any peculiarities to Brazil? \textit{E.g.} How is Brazil different from high-contagion scenarios such as Europe and the USA? How does the quality of the Brazilian health system affect the epidemic? Finally, given what we know so far about the underlying clinical conditions affecting mortality rate, how does Brazil fare? We attempt to provide an answer to these questions in Section \ref{sec:neuralPrediction} by employing publicly-available data alongside a neural regressor. The main conclusions of this work are presented in Section~\ref{sec:conclusions}.

\section{Related Work \label{sec:relatedWork}}

COVID-19's human-to-human transmission is via droplets or by direct contact with an infected person~\cite{lai2020}. An early estimate of the epidemic size in Wuhan, China, was presented in~\cite{wu2020}. The forecast was based on the number of cases exported to international destinations. Several incubation periods have been cited in the literature, namely, 5.2 days~\cite{li2020} to 6.4 days~\cite{backer2020}. Furthermore, estimates of the basic reproduction number $R_{0}$,  a measure describing the average number of secondary cases resulting from an infected person, also vary widely. For example, the intervals $[2.24–3.58]$ and $[1.4–3.8]$ appear in~\cite{lai2020} and~\cite{riou2020}, respectively.


Currently, there are multiple ongoing clinical trials worldwide to assess the effectiveness and safety of certain drugs such as chloroquine, arbidol, remdesivir, and favipiravir~\cite{dong2020}. In vitro data has suggested that chloroquine inhibits virus replication~\cite{wang2020}, although clinical testing has failed to provide such a strong case so far. Also, clinical studies suggest the apparent efficacy of chloroquine phosphate in the  treatment of pneumonia following COVID-19 infection \cite{jianjun2020}. However, as~\cite{touret2020} carefully points out there is a delicate margin between a  therapeutic and a toxic dose. The study reinforces the need for further trials to help validate the claims and design future guidelines.

Given the current lack of proven pharmaceutical solutions, most governments around the world have pursued public policies promoting social distancing, \textit{e.g.:}  closures of schools and universities, remote work when possible, travel restrictions, public gatherings bans, amongst other measures. Additional measures hinge on early detection and isolation, contact tracing, and the use of personal protective equipment~\cite{riou2020}. These measures have been referred to as non-pharmaceutical interventions and a number of studies have been performed in order to assess the effectiveness of these strategies. 

Perhaps some of the best known scientific reports coming out are the COVID-19 series produced by Imperial College. One of these is~\cite{ferguson2020}, which then projected 510,000 deaths in Great Britain and 2.2 million in the United States of America, in the case of an unmitigated epidemic. The authors also projected that even if all patients then these numbers would be revised down to, respectively 250,000 deaths and 1.1-1.2 million. The authors also draw attention to the fact that there is a lag between the introduction of mitigation and the corresponding decrease in hospitalization cases. At the time, their work also strongly emphasized that even for their most optimistic scenario, the number of sick people would far outstrip the available hospital capacity. 

Subsequently,~\cite{flaxman2020} presented estimation of the number of infections and the impact of non-pharmaceutical interventions. This was done by using a semi-mechanistic Bayesian hierarchical model to attempt to infer the impact on 11 european countries. One of their key findings is that the decrease in the number of daily deaths being reported from Italy is in accordance with a significant impact from strict measures introduced weeks beforehand. The authors estimate that (i) between 7 and 43 million individuals, 1.88\% and 11.43\% of the population, to have been infected up to March, 28th; and that (ii) 59,000 deaths had been averted through non-pharmaceutical interventions.

In~\cite{walker2020} the authors analyse different mortality scenarios, from the absence of mitigation measures to policies designed to suppress transmission. They estimate: (i) 7.0 billion infections and 40 million deaths without mitigation; (ii) 4.5 billion infections and 20 millions deaths with mitigation strategies focused on protecting elderly groups and preserving social distancing; (iii) that healthcare systems would be unable to cope even in the latter scenario. Consequently, the work strongly emphasizes the need for public health measures leading to a reduction in transmission rates, in order to avoid the collapse of global health systems.

A recent study proposed a fairly detailed dynamic model to describe the virus spread in China \cite{jia2020}. A drawback is that the model requires 12 parameters that are approximated from real-world data. We argue that such an approximation may lead to highly unreliable estimates given the poor quality and the reliability issues connected to the data made available. Regardless, their main findings, namely that $R_0$ quickly decreases with containment measures and that short quarantines do not suffice to stop the epidemics, hold true and do not depend on the quality of the data.


A simpler SIR (susceptible-infected-recovered) model was applied to data from the UK and Italy \cite{lourenco2020}. The study suggests that (i) the epidemics originated at least a month before the first reported death and (ii) that two to three months of control measures would halt the epidemic. Although the former finding has been used to justify herd immunity strategies, that is hardly in keeping with the reported mortality rates worldwide. To illustrate the point, let us assume a mortality rate of 1\% in the UK. Then, the reported figure of 167 deaths per million as of April 14 (\textit{e.g.,} worldometers.info), would suggest the contagion of approximately 1.67\% of the population. Hence, while the models are useful to guide decision, a holistic and exhaustive analysis is needed to avoid biased assessments.


A model-based analysis aimed at trying to predict mortality rates was described in~\cite{verity2020}. The authors were able to produce age-stratified estimates of the infection fatality ratio. Their findings also estimated the mean duration from symptoms onset to fatality to be 17.8 days, whilst time from symptoms to discharge was calculated as 24.7 days. The overall fatality rate was estimated at 1.38\%. However, older age groups were more afflicted. Fatality increased to 6.4\% among individuals aged 60 or older and reached 13.4\% of those aged 80 or older.


A study compiled and analyzed data from 1099 Chinese patients with confirmed diagnose of COVID-19 \cite{guan2020}. Patients most at risk of: (i) being admitted to an intensive care unit; (ii) requiring ventilator; or (iii) death included people aged 60 or older and also those with coexisting disorders such chronic obstructive pulmonary disease, diabetes, hypertension, coronary heart disease, cerebrovascular conditions, hepatitis B, cancer, chronic renal disease and compromised immune systems. Some authors have also attempted to correlate mortality rates to the Bacillus Calmette-Guérin (BCG) childhood vaccination against tuberculosis~\cite{miller2020}. The authors found a positive significant correlation between the establishment of universal BCG vaccination and mortality rate.

\section{SEIR Model with on/off Strategy \label{sec:SEIRModelOnOffStrategy}}

Following the trend in the literature \citep[e.g.,][]{ferguson2020,flaxman2020}, we used the classical compartmental SEIR (susceptible, exposed, infected, removed) model to describe the virus spread. We argue that, given the uncertainty in the data, a simple but interpretable model can be more useful to provide insights for decision making. The proposed model considers a mean incubation period of 7 days \citep{backer2020} and a mean time to outcome (recovery or death) of 21 days, in line with \citep{verity2020}. We use SEIR instead of the simpler SIR model \citep{lourenco2020} because, in contrast to the latter, it includes the incubation period and allows us to replicate the delayed response to interventions in the system. To model the same spread, \citep{tang19} employed a more detailed approach, whilst \citep{lourenco2020} made use of a simplified SIR model.

\begin{figure}[!ht]
    \centering
    \includegraphics[width=0.48\textwidth]{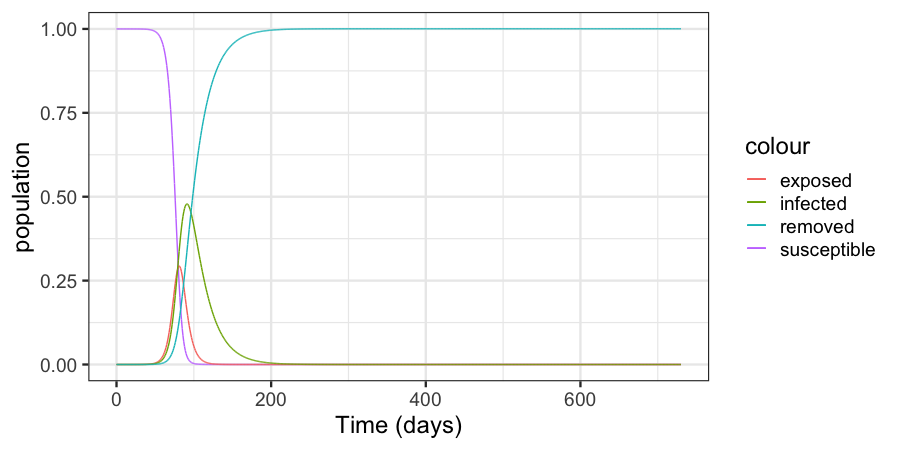}
    \caption{SEIR dynamics for Brazil. \label{fig:original}}
\end{figure}

In order to capture the long-term behaviour, we simulated the system for a period of two years. Figure \ref{fig:original} depicts the dynamics. Notice the steep increase in the infected population, characteristic of the pandemic. Observe also that the proportion of infected individuals peaks around 50\% of the population, which would overload any health system in the world.

The Brazilian health system has experienced a period of decreased investment and counts with around 2 hospital beds per thousand citizens \citep{oecd2020}. To protect this system, some states in the federation are enforcing a lock-down strategy, albeit sometimes challenged by the federal government. This paper proposes a parametric on-off strategy whereby lock-down would be enforced when the number of hospitalisations due to the epidemic  approaches the total number of hospital beds, and removed when the occupation recedes to a lower threshold. For the sake of simulation, we assume a hospitalisation rate of 10\% \citep[e.g.,][]{verity2020}. Hence, the lock-down and and relaxation thresholds can be alternatively set in terms of the total total number of infected patients. Our simulations do not consider the development of curative medication or of an effective vaccine in a two-year horizon. Naturally, should any of these developments occur, the control strategies would have to be completely reformulated.

\begin{figure}[!ht]
    \centering
    \includegraphics[width=0.48\textwidth]{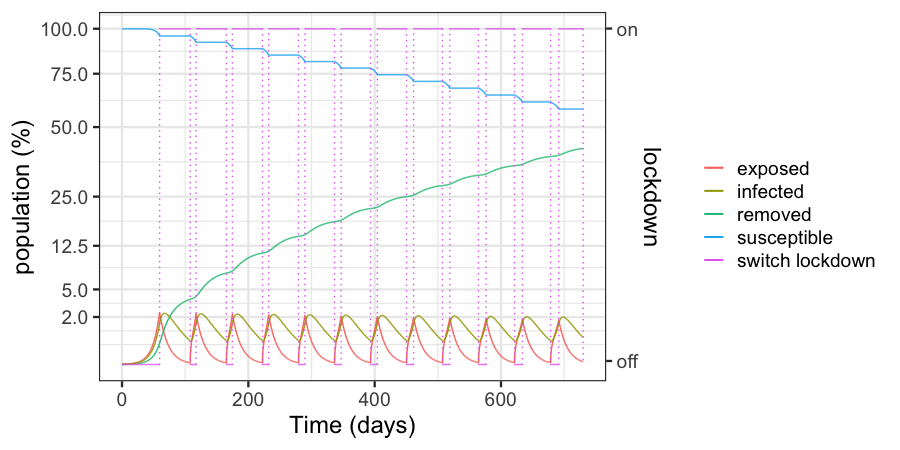}
    \caption{SEIR dynamics for 25\% and 75\% thresholds. \label{fig:25-75}}
\end{figure}

The first strategy is reported in Figure \ref{fig:25-75} and corresponds to activating lock-down whenever the number of infections overcomes 1.5\% of the population, which corresponds to an occupation of 1.5 beds per thousand inhabitants (75\% of the beds). Conversely, the lock-down is relaxed when the bed occupation reaches 25\%, or infection decreases below 0.5\% of the population. Notice that after two years, nearly 40\% of the population will have been infected and therefore be possibly immune. When one considers the results later described in Section \ref{sec:neuralPrediction}, this also means the death of around one to three percent of the population (2.5\% to 7.8\% of the infected population).

\begin{figure}[!ht]
    \centering
    \includegraphics[width=0.5\textwidth]{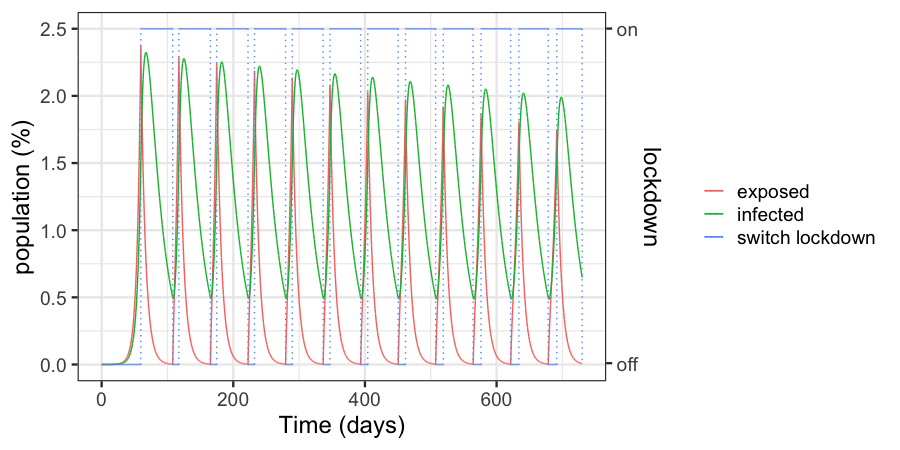}
    \caption{SEIR dynamics for 25\% and 75\% thresholds. \label{fig:25-75zoom}}
\end{figure}

 Figure \ref{fig:25-75zoom} details the evolution of the infected and exposed populations. Observe that, even though the control policy is set for a 1.5\% threshold, the number of infected individuals exceeds 2\% in the peaks because exposed individuals become infected after the onset of the lock-down. Moreover, the peaks decrease over time, as the susceptible population goes down. Notice also that the lock-down periods alternate with comparatively small relaxation intervals. 

\begin{figure}[!ht]
    \centering
    \includegraphics[width=0.5\textwidth]{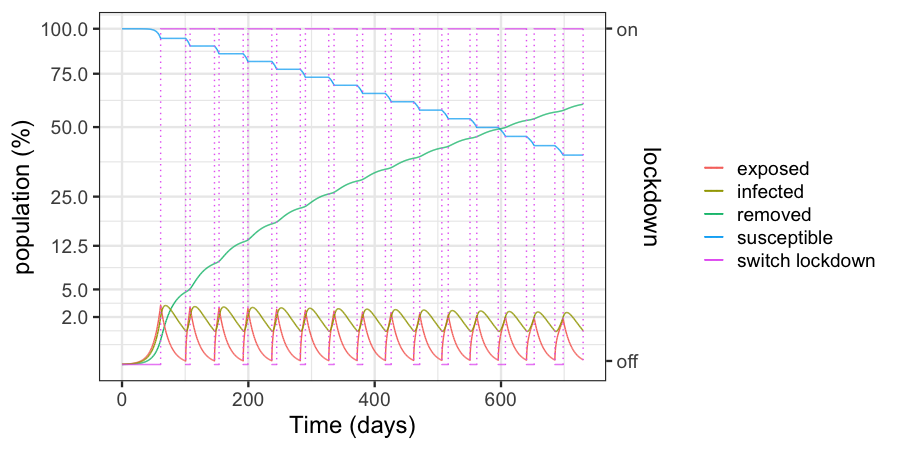}
    \caption{SEIR dynamics for 50\% and 100\% thresholds. \label{fig:50-100}}
\end{figure}

Figure \ref{fig:50-100} depicts the populations for a 50\%-100\% strategy. Lock-down is enforced when hospital beds are full and relaxed when less than half are occupied. With respect to the 25\%-75\% policy, we observe an increase in the infected population, with about 60\% of the population being infected after two years. This is due to the increased occupation in the health system. Unfortunately, in view of the results of that will be presented Section~\ref{sec:neuralPrediction}, the result implies the death of 1.5\% to 4.6\% of the population.

\begin{figure}[!ht]
    \centering
    \includegraphics[scale=0.3]{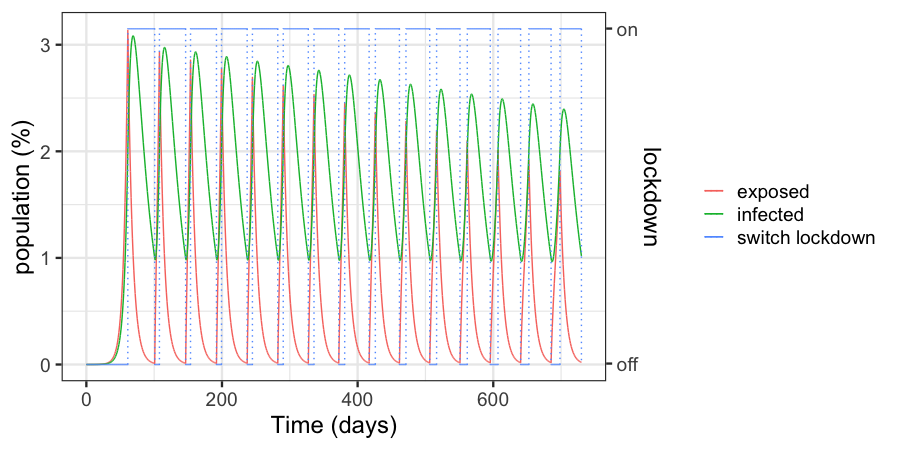}
    \caption{SEIR dynamics for 50\% and 100\% thresholds. \label{fig:50-100zoom}}
\end{figure}

However, as detailed in Fig. \ref{fig:50-100zoom}, the number of required beds is in excess of 3 per thousand inhabitants in the early peaks, signaling that a significant expansion of the health system would be needed. Another insight of the simulations is that the relaxations have to be carefully studied and the thresholds carefully calibrated in order to avoid the collapse of the health system.


\section{Neural Prediction of the Brazilian Case Fatality Rate \label{sec:neuralPrediction}}

The death toll due to COVID-19 in different scenarios is one of the most important quantities to forecast. It can be inferred from the number of infected individuals when the overall case fatality rate (CFR) is known. Unfortunately, the reported Brazilian CFR is not reliable, due in part to insufficient testing~\cite{batista2020b}. Bearing that in mind, we propose a model to predict the Brazilian CFR based on information acquired from COVID-19 data repositories worldwide.

\begin{figure}[htb]
\begin{tikzpicture}[auto,>=latex', transform shape]
\node at (0,0) (textModel1) {Model \#1};
\node at (0,-0.4) (textModel2) {$\vdots$};
\node at (0,-0.95) (textModelN) {Model \#$N$};
\node[draw,scale=0.01] at (0.75,-0.495)   (node0) {};
\node [block, right of=node0, node distance=2cm] (modelSelection) {Model Selection};

\draw[->] (node0) -- (modelSelection);

\node [block, right of=modelSelection, node distance=3.6cm] (PredCommittee) {Prediction Committee};

\draw[->] (modelSelection) -- (PredCommittee);

\node [block, below of=PredCommittee, node distance=1.75cm] (combinationStep) {Combination Step};

\draw[->] (PredCommittee) -- (combinationStep);
\node at (1.7,-2.2) (CFR) {Confidence interval for the CFR};
\draw[->] (combinationStep) -- (CFR);
\draw[->] (modelSelection) -- (CFR);

\end{tikzpicture}
\caption{Block diagram of the proposed committee machine. Note that the combination step is the median operator, and that the confidence interval can be computed using variability statistics derived from the ``Model Selection'' procedure.}
\label{fig:blockDiagramCommittee}
\end{figure}
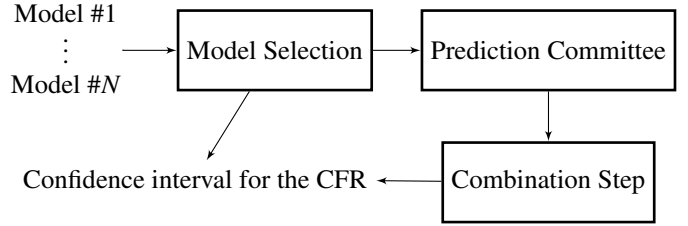

The proposed model utilizes a committee of neural predictors, each with the architecture depicted in Fig. \ref{fig:blockDiagramSingle}. The committee is able to combine individual weak predictors in order to produce an improved overall regression \cite{HaddadBrazilian2018}. Given the variation of the data, and considering the reliability issues surrounding multiple data sources, we use the median of the weak predictors to hedge against outliers \cite{HaddadBrazilian2018,huber2011robust}. Fig. \ref{fig:blockDiagramCommittee} illustrates the committee strategy. 
%
\begin{figure}[htb]
\begin{tikzpicture}[auto,>=latex', transform shape]
\node at (0,0) (textInput) {Input};
\node at (0,-0.4) (textData) {Data};
\node[draw,scale=0] at (0.7,0)   (node0) {};

\node [block, right of=node0, node distance=1.5cm] (Normalization) {Normalization};

\draw[->] (0.7,0) -- (Normalization);

\node [block, right of=Normalization, node distance=3.5cm] (FExtraction) {Features Extraction};

\draw[->] (Normalization) -- (FExtraction);

\node [block, below of=FExtraction, node distance=1.75cm] (NeuralPredictor) {Neural Predictor};

\draw[->] (FExtraction) -- (NeuralPredictor);

\node at (2.5,-1.8) (textCFR){Estimated CFR};

\draw[->] (NeuralPredictor) -- (textCFR);

\end{tikzpicture}
\caption{Block diagram of the architecture of a single neural predictor (\emph{i.e.}, a single model).}
\label{fig:blockDiagramSingle}
\end{figure}
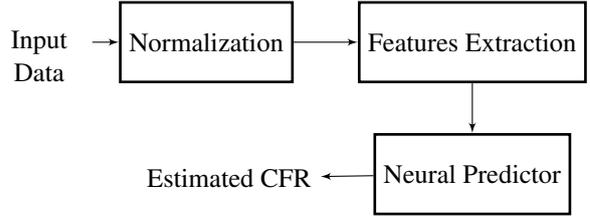

Note that the CFR strongly depends on several risk factors, which can be related to either individual or societal features. Among the former, one finds  chronic medical conditions (especially diabetes~\cite{SinghChloroquine2020}, cardio-cerebrovascular diseases~\cite{BansalCardiovascular2020}, hypertension~\cite{LiClinical2020} and respiratory system diseases~\cite{RodriguezClinical2020}), pregnancy \cite{LiuWhy2020}, obesity~\cite{CrodaCOVID2020} and advanced age~\cite{LiClinical2020}. We can argue that social factors that influence the CFR have attracted less attention, although their impact cannot be dismissed as negligible. Among these factors, one may emphasize: shortage of medical protection in developing countries~\cite{ZhouCovid192020}, risk perception by the community~\cite{BansalClinician2020}, political commitment to allocate resources in order to reduce disaster risks~\cite{LassaMeasuring2019}, disaster risk governance~\cite{OldhamEvolution2018}, appropriate
allocation of humanitarian response and development activity~\cite{GoldschmidtHumanitarian2016}, participatory approaches that change risk management~\cite{BustillosParticipatory2019}, and institutional differences~\cite{EniaRules2016}. It is a challenging task to incorporate such factors in a regression model, mainly due to the absence of reliable metrics for the majority of countries that have experienced COVID-19 dissemination. Fortunately, there are some quantitative features available for the countries of interest that are correlated with the aforementioned factors (\emph{e.g.}, it is expected that the indicator ``Enforcement of regulations'', provided by the Legatum Institute, is correlated with institutional differences between different countries). Overall, a total of ten features were selected, which are detailed in Table \ref{tab:features}. These features are the inputs to the neural predictors. 
Since some countries that present a small number of confirmed COVID-19 cases often have distorted CFRs, the analysis has excluded countries whose number of COVID-19 cases (\emph{i.e.}, variable $x_9$) is lower than 200. After this pruning procedure, 75 countries still remain, resulting in the following matrix of input data $\boldsymbol{X} \in \mathbb{R}^{10 \times 75}$
\begin{equation}
\boldsymbol{X} \triangleq \begin{bmatrix}
\boldsymbol{x}(1) & \boldsymbol{x}(2) & \ldots & \boldsymbol{x}(75)
\end{bmatrix},
\end{equation}
where $\boldsymbol{x}(k) \triangleq \begin{bmatrix}x_1(k) & x_2(k) & \ldots & x_{10}(k) \end{bmatrix}^T$ contains the metrics for the 10 features of the $k$-th country (see Tab. \ref{tab:features}).

{\small
\begin{table}
\begin{tabular}{c|l|c}
Variable & Indicator & Source\\
\hline
$x_1$ & Obesity & LT\\
$x_2$ & Smoking & LT\\
$x_3$ & Healthcare coverage & LT\\
$x_4$ & Raised blood pressure & LT\\
$x_5$ & Public trust in politicians & LT\\
$x_6$ & Enforcement of regulations & LT\\
$x_7$ & Population over age 65 (\%) & WBDI\\
$x_8$ & Fatalities of cardiovascular diseases (\%) & WHO\\
$x_9$ & Number of COVID-19 cases & WOI\\
$x_{10}$ & Number of COVID-19 fatalities & WOI\\
\hline

\end{tabular}
\caption{Input features utilized for the CFR neural regressor. Sources: LT (Legatum Institute), WBDI (World Bank Development Indicators), WHO (World Health Organization Global Estimates 2016), WOI (from \texttt{www.worldometers.info}).}
\label{tab:features}
\end{table}}

\begin{figure}[htb]
\begin{center}
\includegraphics[width=0.4\textwidth]{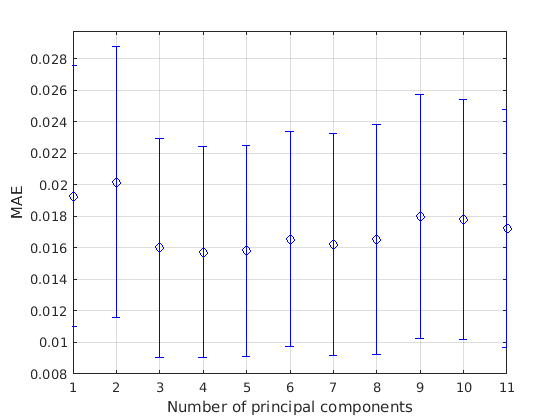}
\end{center}
\caption{Box-plot of the mean absolute error (MAE) obtained with different number of principal components.}
\label{fig:MAE}
\end{figure}

Since the available data is unrealiable, a careful data processing should be performed to guarantee a robust CFR prediction for the Brazilian case. The first processing procedure is executed to enhance the neural network
accuracy (and to speed up training) by reducing the internal co-variate characteristics of the data~\cite{ThakkarBatch2018}. In this first step, each entry of the matrix $\boldsymbol{X}$ is manipulated in order to obtain a normalized matrix $\tilde{\boldsymbol{X}}$, whose elements are computed as
\begin{equation}
\tilde{x}_{i,j} = \frac{x_{i,j}-\hat{\mu}_i}{\hat{\sigma}_i},
\end{equation}
where $\hat{\mu}_i$ (resp. $\hat{\sigma}_i$) is the average (resp. standard deviation) of the $i$-th row of $\boldsymbol{X}$. The chosen neural regressor is the logistic feedforward neuron, whose output, for a set of adjustable parameters $w_i,  \forall i \in \{0,1,\ldots,m-1\},$ is described as
\begin{equation}
y(j) = \frac{1}{1+\text{exp}\left[-w_0-\sum_{i=1}^{m-1}w_i\hat{x}_{i,j}\right]},
\end{equation}
where $\hat{x}_{i,j}$ is distinct from $\tilde{x}_{i,j}$ because of the feature extraction procedure. This procedure is advisable due to lack of sufficient training samples to enforce proper constraints in the neural network parameters, so that the desired estimation is considered a mathematical ill-posed problem~\cite{TorreALeast2012}. It implies that overfitting issues should be mitigated. One tool used for this purpose is the Principal Component Analysis (PCA), which aims to obtain the most compact representation of a high-dimensional data under the sense of least square reconstruction error~\cite{LaiMultilinear2014}. Loosely speaking, it can be described as an unsupervised linear dimensionality reduction technique that presents robust feature extraction properties~\cite{LaiMultilinear2014}. The number $m$ of principal components was selected by $k$-fold cross validation (a kind of model selection technique), in which the data set instances are randomly divided into $k$ disjoint folds with approximately equal size, and every fold
is in turn used to test the model trained from the other $k-1$ folds~\cite{WongDependency2017}. Using $k=10$ folds and training the neural networks with the backpropagation algorithm under the mean quadratic error cost function, the mean absolute error (MAE) for each number of principal components is depicted in Fig. \ref{fig:MAE}.
\begin{table}[htb]
\centering
\begin{tabular}{c|c}
$m$ & Estimated CFR\\
\hline
3 & 0.0213\\
4 & \textbf{0.0255}\\
5 & 0.0266
\end{tabular}
\caption{The estimated Brazilian CFR with respect to the number of principal components. The median estimated is presented in boldface.}
\label{tab:BrazilCFR}
\end{table}

Observe that architectures with three to five principal components perform better. These are selected in our study and provide the estimates in Table \ref{tab:BrazilCFR}. 
%
Observe that the point-wise estimate of the neural committee for the Brazilian CFR is $\overline{\text{CFR}} = 0.0255$. Due to data inaccuracies and to the large differences in the estimated losses in human lives, the variability of such an estimate should be taken into account. In this context, it is more appropriate to adopt a prediction interval, which depends on the variability of the estimator. Since such a variability can be estimated by the $k$-fold cross-validation, one may compute the upper bound $\text{CFR}_{\text{up}}^{\alpha}$ of a confidence interval of $\alpha$\%~\cite{WalpoleProbability2007}. Such an upper bound is  $\text{CFR}_{\text{up}}^{68.27}=0.0548$ (resp. $\text{CFR}_{\text{up}}^{95}=0.0782$) for a confidence interval of 68.27\% (resp. 95\%). 
The median prediction is in line with the official statistics as of April 14 2020 (5.7\% - www.worldometers.info). This suggests that either (i) the underreporting in death cases is similar to the underreporting in the overall cases, or (ii) the testing and reporting biases are captured by the selected variables in the model.

\section{Conclusions \label{sec:conclusions}}


Given the wide assortment of afflictions currently plaguing public available data over COVID-19, it is a challenging task to make reliable predictions concerning the spread and lethality of COVID-19. Consequently, data may be inaccurate and must be utilized with caution, which restricts the reliability of forecasting models constructed with them. It was already demonstrated that an inaccurate confirmed-case data induces nonidentifiability in the model calibrations, which helps to explain the wide range of forecasting variations~\cite{RodaWhyInfectious2020}. For example, underreporting mild cases implies a reduction on the mortality rate~\cite{PetroCOVID2020,BansalClinician2020}. Unfortunately, such inconsistencies in reporting COVID-19 cases are a serious problem, which might sabotage the mitigation of its harmful effects and complicate the outbreak response~\cite{LauInternationallyLost2020}. Additional uncertainties derive from the fact that key characteristics of the transmissibility of COVID-19 (such as whether its transmission can occur before symptom onset) are currently unknown~\cite{HellewellFeasibilityLancet2020}. 

Yet, despite the apparent gaps in knowledge, it is still possible to gain invaluable insight. Namely, by combining the existing SEIR model with an on / off lock-down policies one can see that the impact of the virus will be spread through multiple waves of decreasing amplitude. Such a scenario would effectively mean that there would exist multiple waves requiring flattening over time, in the absence of effective medication, an appropriate vaccine of the development of herd immunity.

Current epidemiological models such as SEIR are also relatively simple. As a result, we developed a neural regressor that considers features that the current literature also deems as important factors in the lethality of COVID-19. This allows for non-linear extrapolations. Again, the issue of data unreliability surfaces. Through careful data processing alongside PCA and k-fold cross validation we believe that it is possible to obtain a more robust CFR prediction for Brazil.

\section*{Declaration of Competing Interest}
We have no conflicts of interest to declare.

\section*{Acknowledgement}

This study was partly supported by the Brazilian Research Council - CNPq, under grants \#431215/2016-2 and \#311075/2018-5 and by Coordenação de Aperfeiçoamento de Pessoal de Nível Superior – Brasil
(CAPES) [Finance Code 001].



\end{document}